\documentclass[twocolumn,aps,showpacs,prl,superscriptaddress]{revtex4}
\usepackage{hyperref}
\usepackage{graphicx}
\usepackage{multirow}
\usepackage{color}

\begin{document}
\title{Engagement in the electoral processes: scaling laws and the role of the political positions}

\author{M. C. Mantovani}
\affiliation{Departamento de F\'isica and National Institute of Science and Technology for Complex Systems, Universidade Estadual de Maring\'a, Av. Colombo 5790, 87020-900, Maring\'a, Paran\'a, Brazil}
\affiliation{Universidade Tecnol\'ogica Federal do Paran\'a, Campus Campo Mour\~ao, 87301-006, Campo Mour\~ao, Paran\'a, Brazil}
\author{H. V. Ribeiro}
\email{hvr@dfi.uem.br}
\author{E. K. Lenzi}
\affiliation{Departamento de F\'isica and National Institute of Science and Technology for Complex Systems, Universidade Estadual de Maring\'a, Av. Colombo 5790, 87020-900, Maring\'a, Paran\'a, Brazil}
\author{S. Picoli Jr.}
\author{R. S. Mendes}
\affiliation{Departamento de F\'isica and National Institute of Science and Technology for Complex Systems, Universidade Estadual de Maring\'a, Av. Colombo 5790, 87020-900, Maring\'a, Paran\'a, Brazil}

\date{\today}

\begin{abstract}
{We report on a statistical analysis of the engagement in the electoral processes of all Brazilian cities by considering the number of party memberships and the number of candidates for mayor and councillor. By investigating the relationships between the number of party members and the population of voters, we have found that the functional form of these relationships are well described by sub-linear power laws (allometric scaling) surrounded by a multiplicative log-normal noise. We have observed that this pattern is quite similar to those previously-reported for the relationships between the number candidates (mayor and councillor) and population of voters [EPL 96, 48001 (2011)], suggesting that similar universal laws may be ruling the engagement in the electoral processes. We also note that the power law exponents display a clear hierarchy, where the more influential is the political position the smaller is the value of the exponent. We have also investigated} the probability distributions of the number of candidates (mayor and councilor), party memberships and voters. The results indicate that the most influential positions are characterized by distributions with very short-tails, while less influential positions display an intermediate power law decay before showing an exponential-like cutoff. We discuss that, in addition to the political power of the position, limitations in the number of available seats can also be connected with this changing of behavior. We further believe that our empirical findings point out to an underrepresentation effect, where the larger city is, the larger are the obstacles for more individuals to become directly engaged in the electoral process.
\end{abstract}



\maketitle

\section{Introduction}
The investigation of social systems is an increasing part of the physicist agenda. In particular, technics and methods from statistical physics have proved to be very useful for unveiling patterns and also for modeling social systems~\cite{Boccaletti,Castellano,Alessandro,Livro,manifesto}. All these investigations are expected to provide a better understanding of the social processes in our society and also to find manners of improve them. A remarkable example are the electoral processes, which have been extensively investigated in the past few years. 

One of the pioneer works (in the statistical physics context) dealing with elections data is due to Costa Filho \textit{et al.}~\cite{Costa, Almeida}, where a power law distribution was proposed to model the number of votes in Brazilian elections. These works were followed by several efforts aiming to model the power law behaviors~\cite{Bernardes,Travieso,Sousa} and also to propose better fits~\cite{Lyra}. However, Fortunato and Castellano~\cite{Fortunato} were the first to investigate the question of universality in the distribution of votes. After analyzing a larger dataset of elections, they concluded that a universal log-normal distribution emerges when considering the number of votes received by the parties. On the other hand, this universally was shown to be not valid for Brazilian elections~\cite{Filho} and, recently, this and other discrepancies were attributed to differences in the election rules among countries~\cite{Chatterjee}. There were also studies showing a tendency of polarization of the voters in mayoral elections~\cite{Araripe,Araujo} and that elections act as a process of information aggregation in which the elected candidates are characterized by a more homogeneous profile~\cite{Tumminello}. The electoral turnout rates~\cite{Bouchaud,Christian,Nadal}, an approach for detecting fraud in elections~\cite{Klimek}, the role of social networks on elections~\cite{Halu}, the evolution of political organizations~\cite{Naim} were also studied.

In spite of elections being currently an effervescent topic, there are still many important questions related to this complex social process that remain without a proper answer. There is no overstatement in affirming that scientists are just beginning to extract quantitive knowledge and also to discover applications for their results in the electoral process. One of these questions is related to the motivations that carry individuals to directly engage in the electoral process by becoming candidate or political party member. { We started to discuss some of these questions in Ref.~\cite{Mantovani}, where we reported that the relationships between the number of candidates (mayor and councilor) and the population of voters are described by power laws surrounded by a multiplicative log-normal noise. Here, we report that the same patterns are observed when investigating the relationships between the number of party memberships and the population of voters. These new findings and those previously-reported thus suggest that the engagement in electoral processes is ruled by similar laws. In addition to this universal behavior, we have observed that the power law exponents display a clear hierarchy where the more influential is the political position the smaller is the value of the power law exponent.} We further have calculated the distributions of the number of candidates (mayor and councilor), number of party memberships and number of voters. The results show that short tail distributions appear for the most influential political positions and that less influential positions display an intermediate power law decay.
%

\section{Data presentation and analysis}

{ We have accessed electoral data of all Brazilian cities made free available by the Brazilian electoral supreme court~\cite{tse} and population of voters also made free available by the Brazilian institute of geography and statistics~\cite{ibge}. The electoral data consist of the number of party memberships from all the 27 Brazilian political parties in each city and for the year 2006, 2008 and 2011. The population of voters was collect to match the years of the data of the number of party memberships. Along the period covering our data, the number of cities have almost not changed (5,564 in 2006 and 5,565 in 2011) and the number of political parties was constant. For shake of convenience, let $V$ denote the population of voters and $N$ be the number of candidates or the number of party memberships. In addition, we have also collected the same Brazilian data used in Ref.~\cite{Mantovani} for investigating the relationships between number of candidates and population of voters.}

Following our previous work~\cite{Mantovani}, we first studied the relationships between $\log_{10} N$ and $\log_{10} V$. Figure~1 shows a scatter plot of {the logarithm of the number of party memberships versus the logarithm of the number of voters.} We note that despite there being fluctuations, a power law tendency is observable. In order to overcome the fluctuations, we have binned the data in $w$ windows equally spaced in $\log_{10} V$ and evaluated the average value of $\log_{10} N$ and $\log_{10} V$ within each window. The circles in Fig.~1 represent these average values where we clearly notice the average power law relationship, that is, the linear fit 
\begin{equation}\label{eq:average} 
 \langle\log_{10} N \rangle = A +\alpha\, \log_{10} V\,
\end{equation}
describes quite well the average values in these log-log scale. It is worth to remark that the values of $A$ and $\alpha$ practically do not change when varying the number of windows. 

\begin{figure}[!ht]
\centering
\includegraphics[scale=0.5]{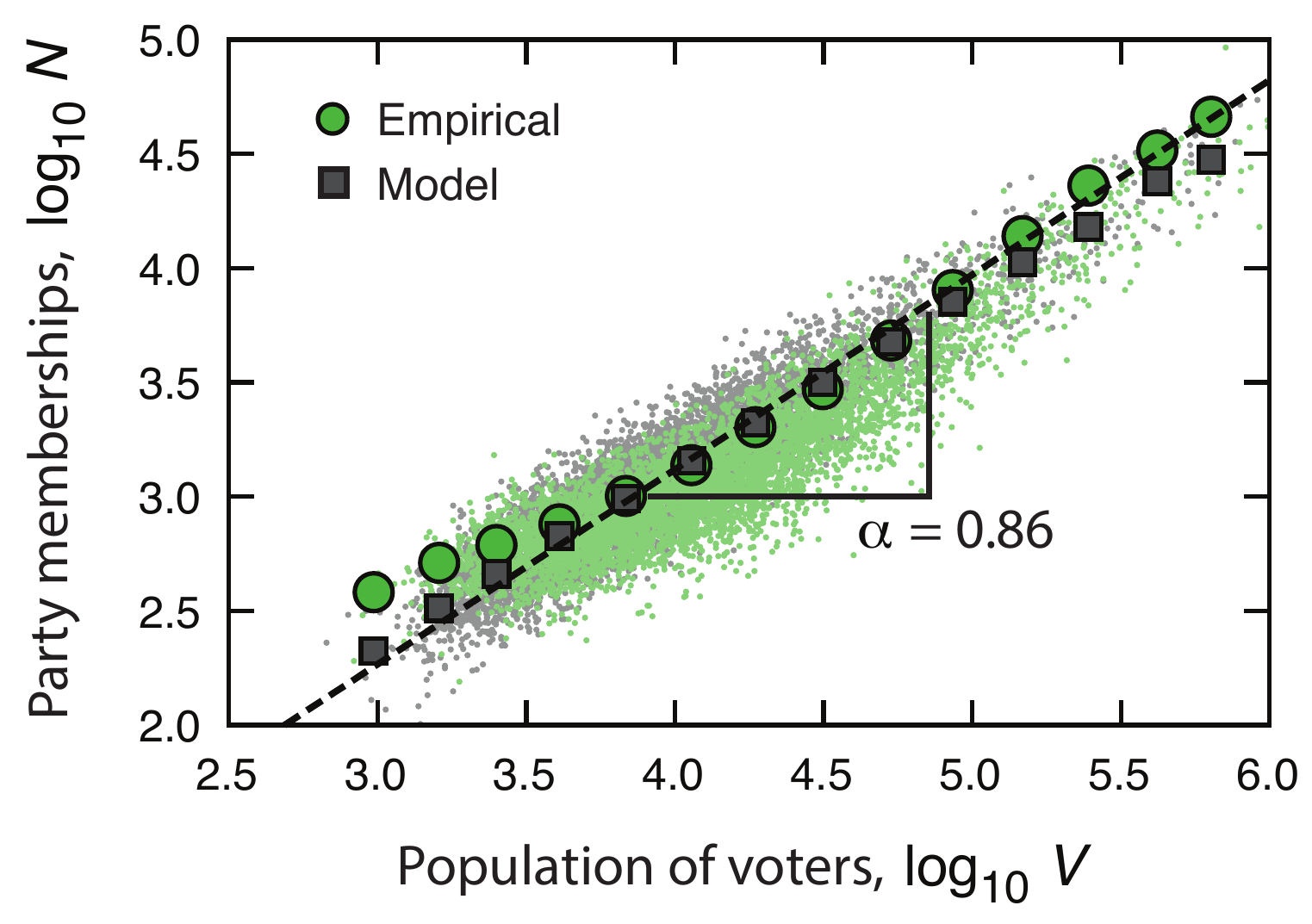}
\caption{
{ (Color online) Scatter plot of the relationships between $\log_{10} N$ (population of voters) and $\log_{10} V$  (number of party memberships) when considering data from the year of 2008. We note that the relationship is approximated by power law tendency surrounded by fluctuations. Here, the green small dots are the values of $\log_{10} N$ and $\log_{10} V$ for each city and the circles are the average values of these points after binning the data in $w$ windows equally spaced in $\log_{10} V$. The dashed line is a linear fit to these average values and the power law exponent is shown in the plot. The simulated results from Eq.~(\ref{eq.model}) are represented by gray dots, which produce the window average values indicated by the gray squares.}
}\label{fig:1}
\end{figure}

In order to check whether values of $\alpha$ {are dependent on time, we have evaluated the average power laws for all years in our database. Figure~2 shows this analysis (that is, average of $\log_{10} N$ versus average of $\log_{10} V$) where we note that the value of $\alpha$ practically does not chance in time. In particular, we have found $\alpha=0.87\pm0.02$, $\alpha=0.86\pm0.02$ and $\alpha=0.85\pm0.02$ respectively for the years 2006, 2008 and 2011. Thus, we cannot reject the hypothesis that the values of $\alpha$ are constant over time after taking the standard errors of $\alpha$ into account.}


\begin{figure}[!ht]
\centering
\includegraphics[scale=0.5]{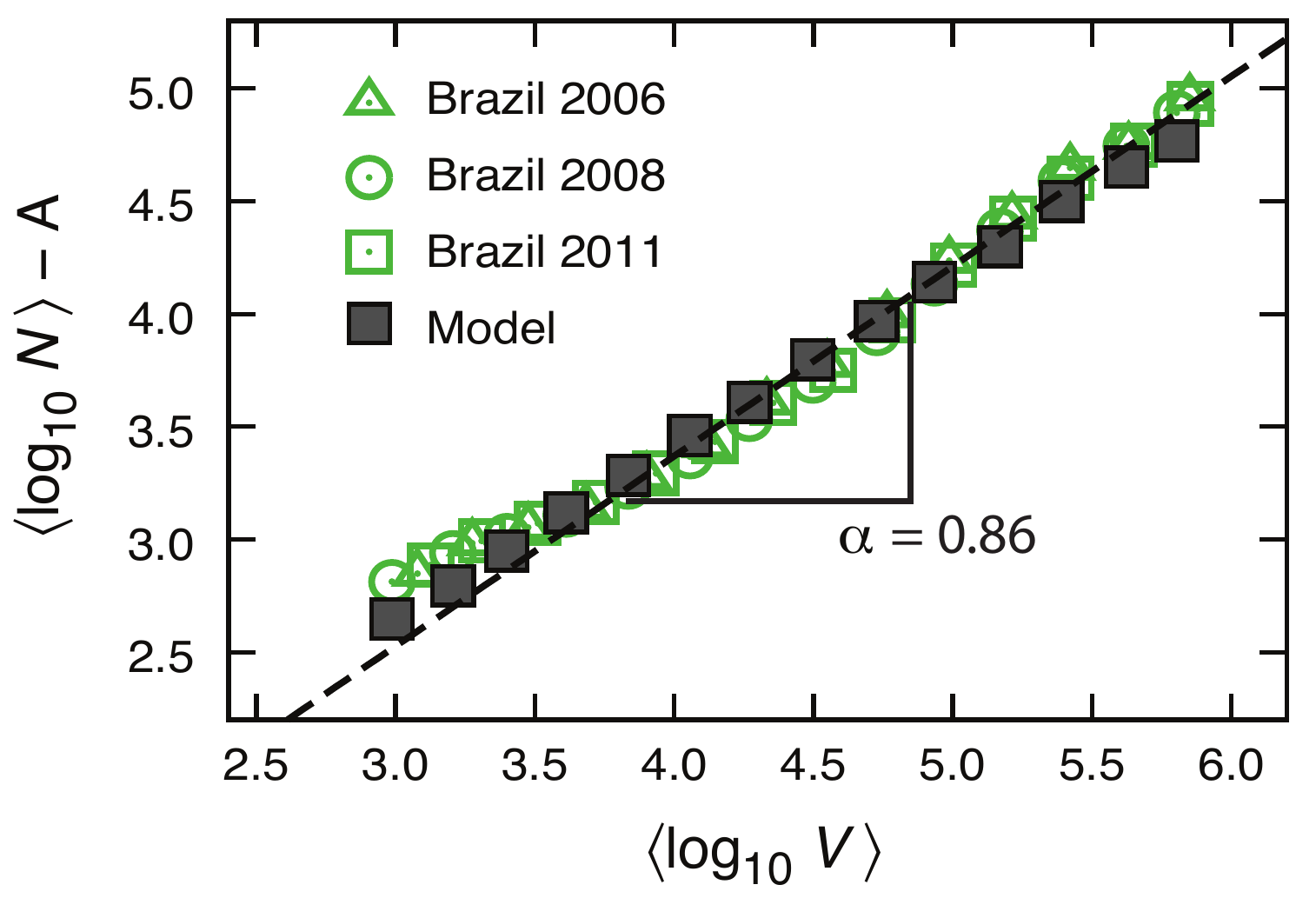}
\caption{
{(Color online) Average power law relationships between $\log_{10} N$ (population of voters) and $\log_{10} V$ (number of party memberships). Note that we have analyzed all the years in our database and the values of $\alpha$ (the slope of these curves) practically do not change in time. The average value of $\alpha$ is shown in the plot and the dashed line represents a power law with this exponent. The gray squares represent the simulated results from Eq.~(\ref{eq.model}).}
}\label{fig:2}
\end{figure}

In addition to the average values, we have also studied the fluctuations around the power law tendencies. We have calculated the variance $\sigma^2$ of $\log_{10} N$ within each window as a function of average value of $\log_{10} V$. Figure~3(a) shows that these variances are almost a constant function of $\log_{10} V$ and that there is no clear temporal evolution for their average values. The average value of the variance over $\log_{10} V$ and over the years is $0.027\pm0.002$.

{Still following our previous work, }we have also investigated the distributions of the fluctuations around the power law tendencies by considering the normalized residuals
\begin{equation}\label{eq:noise}
\xi(V)=\frac{\log_{10} N(V) - \langle\log_{10} N \rangle}{\sigma}\,,
\end{equation}
where $\sigma$ is the standard deviation of $\log_{10} N$ evaluated within the windows. Figure~3(b) shows the distributions of $\xi$ where we have found that the Gaussian distribution with zero mean and unitary variance describes quite well the empirical distributions for all the data. In particular, after applying the Pearson chi-squared hypothesis test, we have concluded that normality cannot be rejected {for all years}. 

\begin{figure}[!ht]
\centering
\includegraphics[scale=0.5]{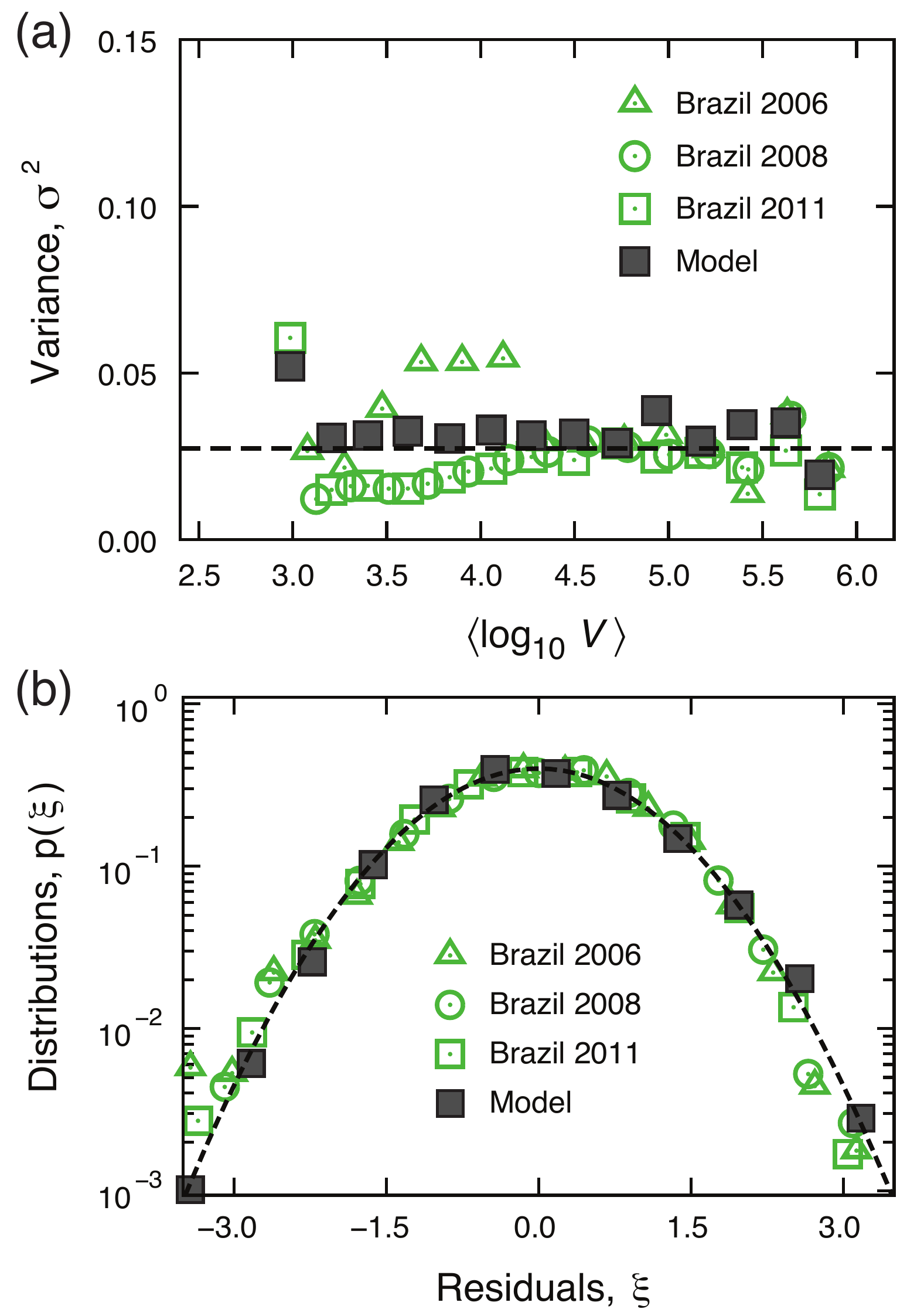}
\caption{
{(Color online) Fluctuations around the power law tendencies. (a) Variance $\sigma^2$ of the values of $\log_{10} N$ versus $\langle\log_{10} V \rangle$ evaluated after binning the data in $w$ windows equally spaced in $\log_{10} V$. We note the almost constant value of $\sigma^2$ over the population of voters. The horizontal dashed line is the average value of $\sigma^2$ over $\log_{10} V$ and over the years ($0.027\pm0.002$). (b) Distribution of the normalized residuals ($\xi$) around the power law tendencies. The dashed line is the Gaussian distribution with zero mean and unitary variance. In both plots, the gray squares represent the simulated results from Eq.~(\ref{eq.model}).}
}\label{fig:3}
\end{figure}

{Based on the previous results, {we thus concluded } the patterns exhibit by relationships between the number of party memberships and the population of voters are quite similar to those we have reported for the relationships between number of candidates (mayor and councilor) and population of voters~\cite{Mantovani}. Furthermore, as we have previously proposed, all these empirical findings can be modeled by the} following stochastic-like equation
\begin{equation}\label{eq.model}
N = \mathcal A \,\eta(V) V^{\alpha}\,,
\end{equation}
where $\mathcal A$ is a constant, $\eta(V)$ is a stochastic noise and $\alpha$ is the power law exponent of the relationship between $N$ and $V$. Notice that if we apply the logarithm in both sides of Eq.~(\ref{eq.model}) and compare the results with the Eqs.~(\ref{eq:average}) and (\ref{eq:noise}), we will figure out that $\log_{10} \mathcal A = A$, $\log_{10} \eta(V) = \sigma\, \xi(V)$ and $\langle\xi(V) \rangle=0$. We further observe that since $\xi(V)$ is normally distributed, $\eta(V)$ must be distributed according to a lognormal distribution. Thus, Eq.~(\ref{eq.model}) represents a multiplicative stochastic process driven by a lognormal noise. An analogous behavior was also reported for scaling laws in crime data~\cite{Bettencourt}, which reinforces the universally of our empirical finding. We have simulated the Eq.~(\ref{eq.model}) by considering the average values of $\alpha$ and $\sigma$. The simulation results are shown in Figs.~1, 2 and 3 by gray squares. We note that Eq.~\ref{eq.model} describes quite well all the empirical findings we have reported.

\section{The role of the political position}

\begin{figure}[!ht]
\centering
\includegraphics[scale=0.5]{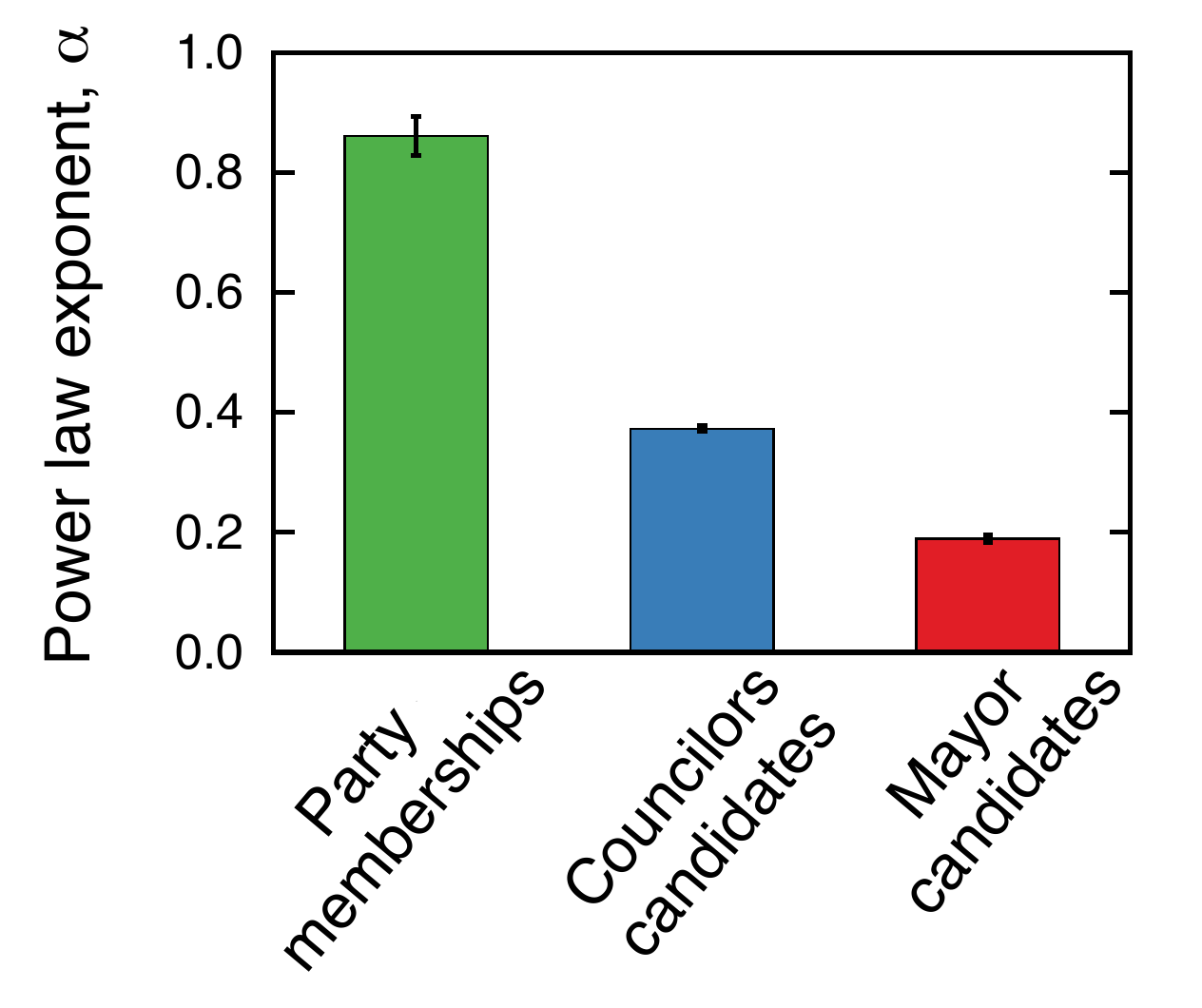}
\caption{
{(Color online) The role of the political positions. Average values of the power law exponents $\alpha$ for each political position. The exponents of the relationships between number of councilor and mayor candidates and the population of voters were obtained from Ref.~\cite{Mantovani}. }We note that the average of $\alpha$ is smaller for the more influential political positions. The error bars are standard deviations from the mean value of $\alpha$.
}\label{fig:4}
\end{figure}

We have thus identified patterns in the engagement in the electoral processes that seem to hold for different political positions. From number of party memberships {(reported here)} to number of councilor and mayor candidates {(reported in Ref.~\cite{Mantovani}), the} empirical findings have shown that a robust average scaling law with the population of voters emerges and that the fluctuations surrounding this average tendency have an almost constant variance and a log-normal distribution. However, within this universal behavior we {can also verify} that the power law exponents $\alpha$ enabled us to distinguish the different political positions. {By comparing the average value of the power law exponent reported here for the number of party memberships to those previously-reported for councilors and mayor candidates, we observe that there is a clear hierarchy for the values of $\alpha$, as shown in Fig.~\ref{fig:4}. Moreover, we also note that the more influential is the political position, the smaller is the value of $\alpha$.} This result may connected to the number of available seats for each of these political positions: there is only one mayor in each city, 9 councilors in cities with less than 47,600 inhabitants and, in principle, the only limitation for the number of party memberships of a city is its population of voters. The case of party memberships is thus very intriguing, since in the lack of restrictions we could expect $N$ to be linear with the population of voters $V$. {In addition to number of seats, the fact that some parties are not represented in small towns may help to limit the number of candidates. This limitation however should affect more the number of mayor candidates, since it is not possible to have more than one mayor candidate from the same party in a given city; meanwhile, there is no such limitation for councilor candidates and for party memberships. The case of party memberships is not even conditioned to presence of parties in a city. Thus,} the intrinsic sub-linear behavior and the fact that $\alpha$ becomes much smaller than one for the most influential political positions may be in somehow connected to a kind of underrepresentation effect, that is, larger cities may not provide the necessary conditions for more individuals to become directly engaged in the electoral process.

\begin{figure}[!ht]
\centering
\includegraphics[scale=0.5]{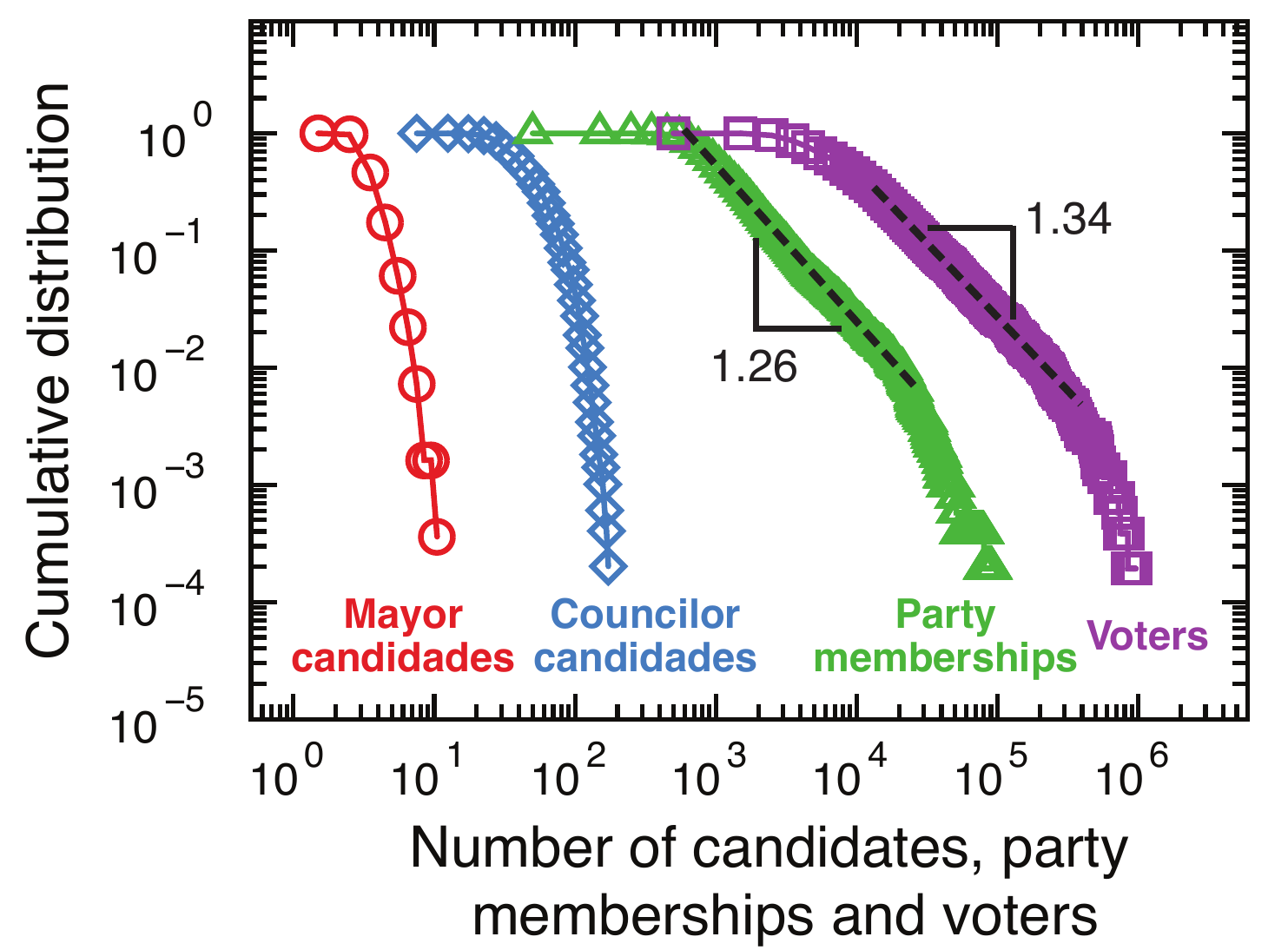}
\caption{
(Color online) Cumulative distributions of the number of mayor candidates (circles), councilor candidates (diamonds), party memberships (triangles), and population of voters (squares) when considering data from the year 2008 (very similar distributions are obtained for the other years). We note the shift of the distributions with the increasing in the number of available seats. We further observe that the distributions for mayor and councilor are characterized by exponential asymptotic behavior, while the distributions for party memberships and voters display an approximate power law behavior before showing an exponential-like cutoff. The dashed lines show the regions where the distributions can be approximated by power laws (the exponents are shown in the plot).
}\label{fig:5}
\end{figure}

Another intriguing behavior is observed when we calculate the probability distributions of the numbers of mayor candidates, councilor candidates, party memberships, and voters. Figure~\ref{fig:5} shows the cumulative distributions of these numbers. We note that for mayor and councilor candidates the distributions are characterized by short tails that can be approximated by exponentials decays. On the other hand, for party memberships and voters the distributions exhibit an intermediate behavior that can be approximated by power law decays (dashed lines in Fig.~\ref{fig:5}) before showing an exponential-like cutoff. This changing of behavior may be connected to the limitation in the number of available seats, where exponentials decays appear for positions with limited number of seats (mayor and councilor) and approximate power law decays appear for positions in which there is no clear limitations in the number of seats (party memberships and voters). Actually, because voting is compulsory in Brazil, the number of voters can be thought as the fraction of population that is older than 18 years and, in fact, we have verified that number of voters increases linearly with the population size. 

\section{Summary and conclusions}
In this paper, we have studied the engagement in the electoral process in all Brazilian cities by taking into account the number of {party memberships and the number of candidates for mayor and councillor. We have shown that an average power law relationship surrounded by a multiplicative log-normal noise emerges when investigating the relationships between the number of party memberships and population of voters. We also observed that the same behavior was previously-reported for the relationships between the number candidates (mayor and councilor) and population of voters.} When comparing the value of the power law exponents, we have found a clear hierarchy of values where the most influential positions display the smaller values for the power law exponents. We have also evaluated the probability distributions of the number of candidates (mayor and councilor), party memberships and voters. The results have pointed out that the most influential positions (mayor and councilor) are characterized by distributions with very short-tails, while less influential positions (party memberships) display an intermediate power law decay. We have argued that, in addition to the political power of the position, limitations in the number of available seats may also be connected with this changing of behavior. We further believe that the strong sub-linear behavior and the exponential-like cutoff provide clues pointing out to an underrepresentation effect, where the larger city is, the larger are the obstacles for more individuals to become directly engaged in the electoral process. 

\acknowledgements
We thank Capes, CNPq and Funda\c{c}\~ao Arauc\'aria for financial support. HVR is especially grateful to Funda\c{c}\~ao Arauc\'aria for financial support under grant \mbox{number 113/2013}.

\end{document}